\documentclass[aps,epsfig]{revtex4}
\usepackage{graphicx}
\begin{document}
\def\be{\begin{equation}}
\def\ee{\end{equation}}
\def\bfi{\begin{figure}}
\def\efi{\end{figure}}
\def\bea{\begin{eqnarray}}
\def\eea{\end{eqnarray}}

\title{Heat exchanges in coarsening systems}

\author{Federico Corberi}
\affiliation {Dipartimento di Fisica ``E.R. Caianiello'', 
Universit\`a di Salerno, 
via Ponte don Melillo, 84084 Fisciano (SA), Italy.}
\author{Giuseppe Gonnella}
\affiliation {Dipartimento di Fisica, Universit\`a di
Bari {\rm and} Istituto Nazionale di Fisica Nucleare, Sezione di  Bari, via Amendola 173,
70126 Bari, Italy}
\author{Antonio Piscitelli}
\affiliation {Dipartimento di Fisica, Universit\`a di
Bari {\rm and} Istituto Nazionale di Fisica Nucleare, Sezione di  Bari, 
via Amendola 173,
70126 Bari, Italy}

\begin{abstract} 

This paper is a contribution to the understanding of the thermal properties
of aging systems where statistically independent 
degrees of freedom with largely separated timescales
are expected to coexist. Focusing on the prototypical  case of 
quenched ferromagnets, where fast and slow modes can be respectively associated
to fluctuations in the bulk of the coarsening domains and to their interfaces,
we perform a set of numerical experiments specifically designed 
to compute the heat exchanges between different degrees of freedom. 
Our studies promote a scenario with fast modes
acting as an equilibrium reservoir to which interfaces may release heat
through a mechanism 
that allows fast and slow degrees to maintain
their statistical properties independent.

\end{abstract}
\maketitle

PACS: 

\section{Introduction} \label{Heat}

When two equilibrium systems in contact with reservoirs at
different temperatures $T_1$ and $T_2$ are put in contact, heat flows
from the hotter to the colder
and the usual equilibrium concept of 
temperature can be used to establish the average direction of heat fluxes. 
Deviations from the average are described by a fluctuation principle
known as the Gallavotti-Cohen relation \cite{GC}, namely
\be
\ln \frac {P(Q_{1,2})}{P(-Q_{1,2})}=Q_{1,2} \left (\frac{1}{T_2}-\frac{1}{T_1} \right ),
\label{fluc} 
\ee 
where $P(Q_{1,2})$ is the probability that the first system exchanges
the heat $Q_{1,2}$ in a certain (sufficiently long) time with the other one
and we have set to unity the Boltzmann constant. 

In an attempt to generalize ideas of equilibrium thermodynamics 
to non-equilibrium systems, an {\it effective temperature} $T_{eff}$ has been
introduced in \cite{7dipeliti} in the context of non-equilibrium stationary systems
and in \cite{89dipeliti} for aging systems.
In this formulation, $T_{eff}$ can be evinced from
the relation  
\be
R(t,t_w)=T_{eff}(t,t_w)^{-1}\frac{\partial C(t,t_w)}{\partial t_w}
\label{efftemp}
\ee
between the auto-correlation function $C(t,t_w)=\langle O(t)O(t_w)\rangle
-\langle O(t)\rangle \langle O(t_w)\rangle$ of the observable
$O$ at times $t$ and $t_w<t$ and the associated linear response function,
$R (t,t_w)= \lim _{h\to 0}\delta O(t)/\delta h(t_w)$,
where $h$ is a small perturbation introducing an extra term $-h(t)O$ in the
Hamiltonian. In equilibrium states, due to the fluctuation-dissipation 
theorem, $T_{eff}=T$ is the usual temperature which is 
independent of the two times and on the chosen observable $O$.
Out of equilibrium, in principle $T_{eff}$ may depend on $t,t_w$ and $O$
and is not necessarily related to the temperature of the reservoir.
In \cite{peliti} it was argued that $T_{eff}$ exhibits the
properties of a temperature, in the sense that it regulates
the direction of heat fluxes favouring thermalization.
Specifically, if different degrees of freedom {\it effectively}
interact on a given time scale, heat flows from those with 
higher $T_{eff}$ to those with a lower one.  
The concept of effective temperature is possibly relevant when thermal
flows are small; in the context of aging systems this usually
amounts to consider the large time 
regime. Generally, in this asymptotic domain one can identify
fast and slow degrees of freedom  
whose typical timescales $t_{f}$ and $t_{s}$ become widely separated.
Indeed, while $t_f$ is generally age-independent, $t_s$ is usually
an increasing function of the age $t_w$ of the sample. 
When this happens, one can probe fast or slow degrees increasing the times
$t,t_w$ along particular directions in the 
$t,t_w$ plane. Specifically, by letting $t_w \to \infty$ keeping
$t-t_w$ finite, since $t-t_w\ll t_{s}$ the slow degrees act
adiabatically and can be considered as static. In this
time sector, denoted as the short time difference (or quasi-equilibrium) regime, 
one probes the dynamical behavior of fast modes. Conversely,
letting again $t_w \to \infty$ but keeping $t/t_w$ 
(or some different combination of $t$ and $t_w$ in specific systems)
finite one has $t-t_w \sim t_w \sim t_{s}$, focusing the analysis on the timescale
of the slow (or aging) components. In this time sector, usually
denoted as aging regime, the fast degrees
act as a stationary background and the time evolution of
the slow modes is probed. 
In a class of aging systems, among which mean field p-spin models
and coarsening systems \cite{parisi}, the effective temperature 
(here and in the following we simply quote the name for the quantity
$T_{eff}$ defined through Eq. (\ref{efftemp}), without necessarily conforming
to the debated interpretation of it as a thermodynamic temperature) 
takes two
different well defined values in the short time and in the aging
regime. In the former sector one has $T_{eff}=T$, the bath temperature, while 
in the aging regime $T_{eff}=T_s>T$. A major difference between
mean field glassy models and phase-ordering systems is that $T_s$
is finite in the former, while it is infinite in the latter.
This simple two-temperature scenario,
as opposed to that of a continuously varying $T_{eff}$ displayed
by models with full replica symmetry breaking, such as the 
Sherrington-Kirkpatrik spin glass, or some systems 
without local detailed balance \cite{detbal},
is amenable of a simple 
interpretation 
in terms of fast and slow components:
The former which are already at the equilibrium temperature and
the latter retaining a self-generated higher temperature $T_s$.
Generally, the existence of different effective temperatures
in the asymptotic domain implies that thermalization among different
degrees of freedom is suppressed. While on the whole the mechanism
whereby thermalization is avoided is uncertain,
in the case of coarsening systems, at least in the
solvable large-$N$ model ($N$ being the number of components of the
vector order parameter) 
fast and slow degrees
can be explicitly exhibited and their
statistical independence can be proved \cite{noilargen}.
This is consistent with two temperatures being sustained and coexist
in the system. However, the issue of heat exchanges occurring between
fast and slow modes has never carefully been analyzed so far.

In this paper we tackle this question numerically in 
the physical case of coarsening systems with $N=1$. 
We check to some extent the
picture with slow and fast degrees, studying the heat exchanges
between them, and suggesting a mechanism whereby 
statistical independence is preserved and
effective thermalization does not occur. 
In particular, we will consider
the two-dimensional Ising model quenched below $T_c$.
This is perhaps the simplest
case where the two-degrees/two-temperature scenario can be investigated.
We mention that the Ising model in $d=1$ exhibits an anomalous
coarsening behavior with a continuously varying $T_{eff}$ \cite{lippiello} 
which, moreover, is found to depend on the particular observables $O$
entering $R$ and $C$ in the definition. These
features, which are presumably related to the fact that the model
is at the lower critical dimensionality $d_L$ \cite{claudio},
make the interpretation of $T_{eff}$ as a genuine temperature
doubtful. However, for $d>d_L$ these features disappear and 
the two-temperature scenario is more sound.

The paper is organized in five sections. In Sec. \ref{single} we
review the kinetics  of binary systems in contact with a single thermal
bath. In Sec. \ref{two} we consider a system in contact with two reservoirs 
and introduce the method to compute heat exchanges. In Sec. \ref{Numsetup} we describe
our numerical experiments to study heat flows in phase-ordering
systems.
In Sec. \ref{conclusions} we draw the conclusions and discuss the perspective of this work.

\section{Dynamics of binary systems in contact with a single heat bath} \label{single}

The dynamics of a non-disordered binary system (i.e. 
one that can be described by the ferromagnetic Ising model) in contact with
a thermal bath, 
has been extensively studied since a long time.
Although exact results are scarce, the overall phenomenology
is nowadays quite well understood \cite{bray}.
Hereafter we recall some basic facts 
that will be needed in the following and
specify our dynamical model, that will be later (in Sec. \ref{two}) generalized
to the case of a system in contact with two reservoirs.

In the high temperature disordered phase, for $T>T_c$, 
the system is ergodic and equilibrium is achieved
on microscopic timescales (if not too close to $T_c$) starting from any configuration.
On the other hand, the evolution below $T_c$ depends strongly on the initial state.
In particular, if the sample is prepared in a broken symmetry 
configuration (for instance, with all spins up) the dynamics
converges on a microscopic time to the broken symmetry equilibrium
state. This is characterized by finite coherence length $\xi $ and
relaxation time $t_{eq}\propto \xi ^z_{eq}$, where $z_{eq}$ is the usual
equilibrium dynamical exponent. $\xi$ and $t_{eq}$
can be interpreted as the spatial extent 
and the persistence of thermal islands of reversed spins 
in the sea of opposite magnetization. These fluctuations will be
denoted as {\it fast}, because $t_{eq}$ is finite (and small
if $T$ is sufficiently far from $T_c$).
Conversely,
starting from a typical high temperature equilibrium configuration, namely a disordered one, 
and quenching to a sub-critical temperature, a coarsening dynamics sets in where domains of the two
magnetized phases with a typical size $L(t)$ grow in time.
For the case of a dynamics with non-conserved order parameter
considered in this paper one has $L(t)\propto t^{1/z}$, with $z=2$,
for sufficiently long times. If an interface is present at time $t_w$
in some part of the system, 
it must move a distance of order $L(t_w)$ in order to produce an effective decorrelation. 
Since this takes a time of order
$t_{s}\propto L(t_w)^z \propto t_w$, the characteristic decorrelation time
due to interface motion 
is $t_{s}\propto t_w$. This is the essence of the aging phenomenon,
where the evolution gets slower as time elapses and $t_{s}$ diverges.
The interfacial degrees of freedom responsible for the aging process will be 
denoted as {\it slow}.  
However, these are not the only degrees contributing to the dynamics.
Indeed, the interior of domains is basically in equilibrium in one of the
broken symmetry phases described above. It is then
characterized by the fast thermal fluctuations with
characteristic timescale $t_f=t_{eq}$. 
In conclusion, in the late stage evolution of a coarsening system
there are two dynamical components, fast and slow, whose timescales
are widely separated in the large $t_w$ limit. 

In this paper we will consider the two-dimensional Ising model, where 
$N$ discrete spin variables $\sigma _i=\pm 1$ defined on the sites $i$ of 
a square lattice are governed by the Hamiltonian
$H(\sigma )=-J\sum _{\langle i,j\rangle}\sigma _i \sigma _j$.
Here $\sigma $ denotes the whole spin configuration and
the sum is running over all nearest neighbours couples $\langle i,j \rangle$.
The critical temperature of the model is $T_c=2J[atanh (1/\sqrt 2)]^{-1}\simeq 2.269 \, J$.
In the following we will set $J=1$.

We introduce a dynamics where a single spin
update is attempted in each timestep;
time will be measured in Montecarlo steps (or sweeps), corresponding
to $N$ timesteps. Metropolis single spin transition rates
\be
w(\sigma)=min\left [\exp \left ( \frac{-\delta E}{T}\right ),1\right ], 
\label{metrop}
\ee
where $\delta E$ is the energy change due to the flip, will be used. 

In order to study the evolution, let us
consider the energy (per spin) $E(t)=(1/N)\langle H(t)\rangle$
whose equilibrium value is $E_{eq}$. Here $\langle \ldots \rangle$
means an average over thermal noise and initial conditions.
Since the excess energy $\Delta E(t)=E(t)-E_{eq}$ is associated to the
domain boundaries, in the coarsening stage one has 
\be
\Delta E(t)\propto L(t)^{-1}\propto t^{-1/2}.
\label{eq:E}
\ee
This behavior lasts until $L(t)$ becomes comparable with the
size of the system and the system either equilibrates or remains
trapped (particularly at low temperatures) in long lived metastable 
states with one (or few) spanning interfaces. 
If the thermodynamic limit is taken from the beginning 
the system is free from finite size effects and
coarsening lasts forever: Equilibration is then never observed.

\begin{figure}
    \centering
    
   \rotatebox{0}{\resizebox{.45\textwidth}{!}{\includegraphics{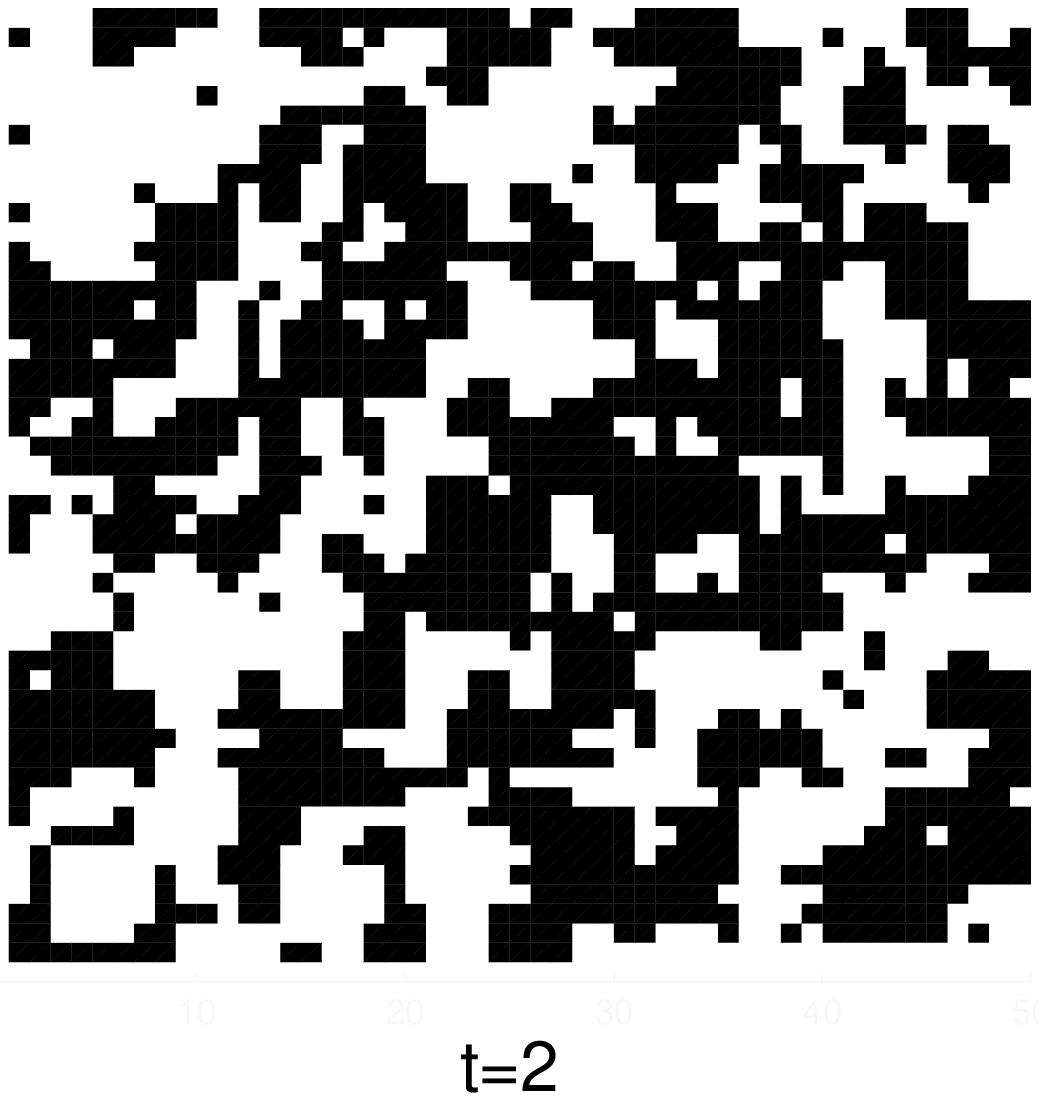}}}
   \rotatebox{0}{\resizebox{.45\textwidth}{!}{\includegraphics{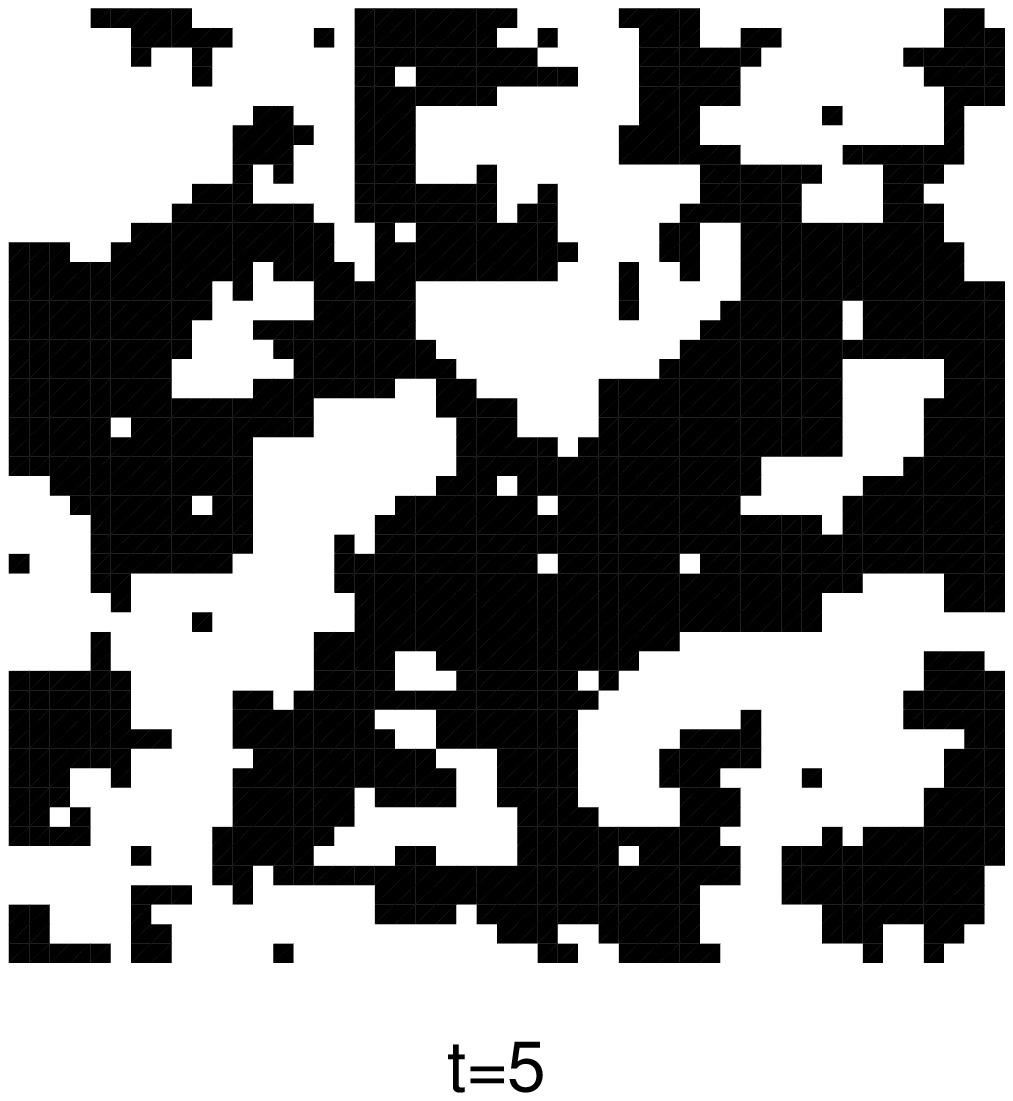}}}
   \rotatebox{0}{\resizebox{.45\textwidth}{!}{\includegraphics{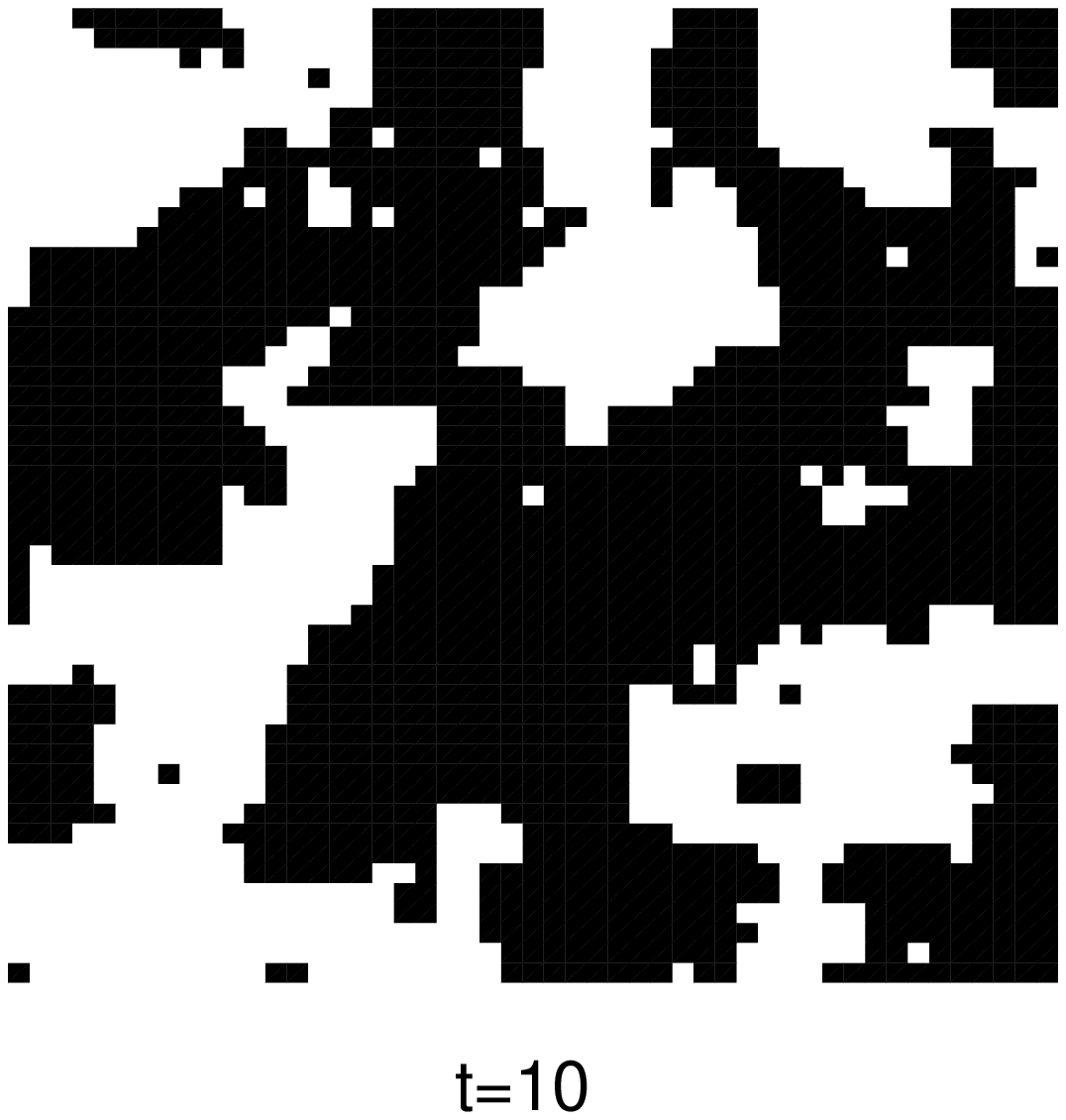}}}
   \rotatebox{0}{\resizebox{.45\textwidth}{!}{\includegraphics{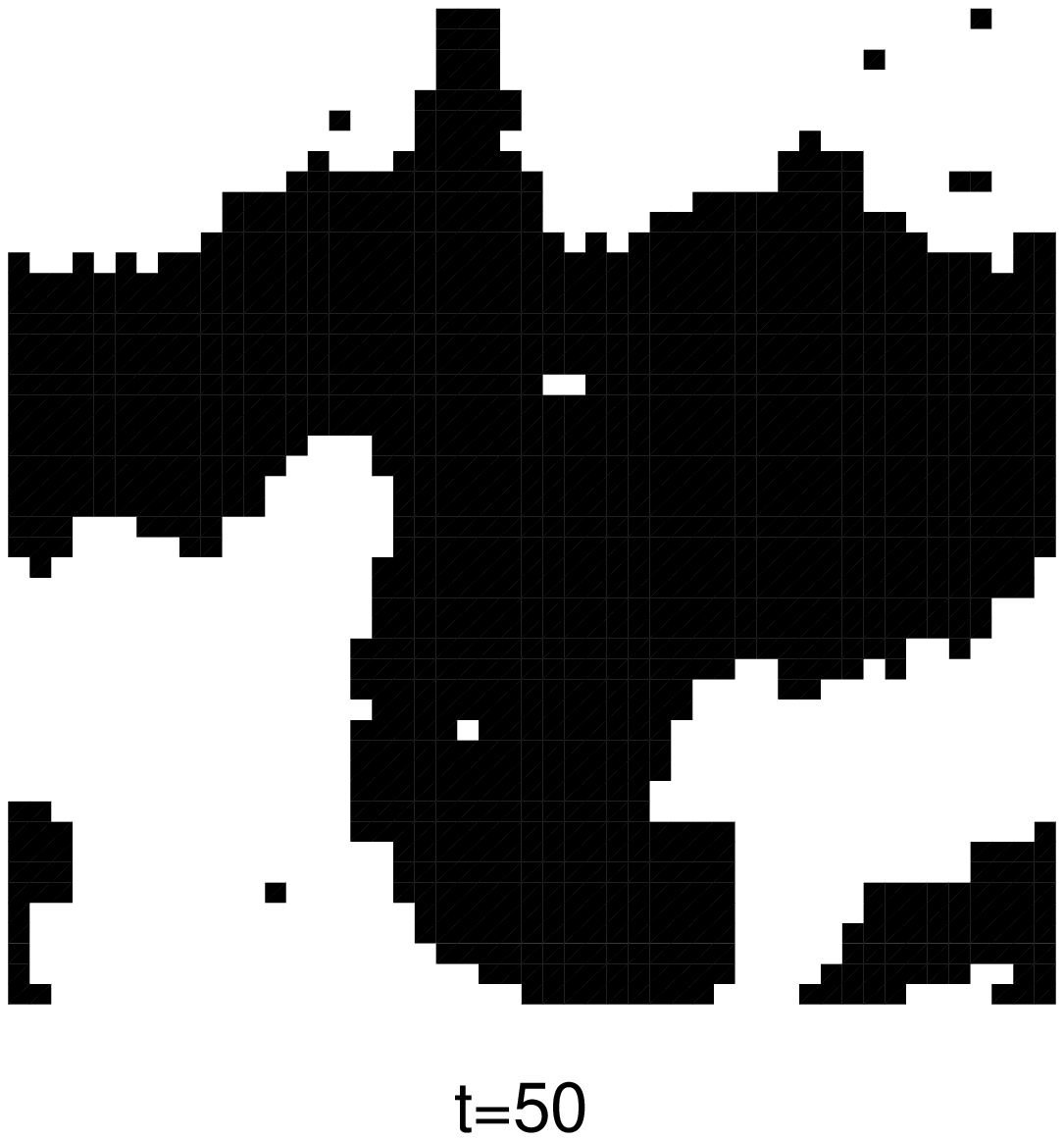}}}
    \caption{(Color online). Domain configurations at different times
    $t$ after a quench to $T=1.5$. Black and white regions
    correspond to up and down spins. The system size is $N=50$x$50$.}

\label{figsnap}
\end{figure}

Snapshots of the configurations of the system 
after a quench from a disordered state to a 
subcritical temperature $T<T_c$ are shown in Fig. (\ref{figsnap}), where 
the growing domains are visible. 
One can also see small regions of reversed spins
inside the growing domains. These are the fast degrees of freedom,
with the equilibrium behavior discussed above.

\begin{figure}
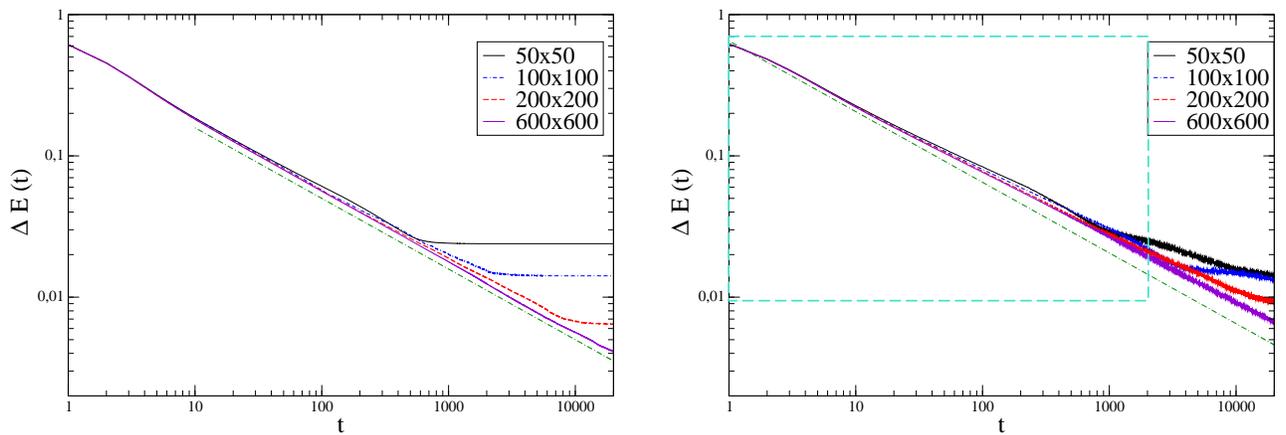

    \centering
    
   \rotatebox{0}{\resizebox{.45\textwidth}{!}{\includegraphics{figura_2a.eps}}}
   \hspace{.5cm}
   \rotatebox{0}{\resizebox{.45\textwidth}{!}{\includegraphics{figura_2b.eps}}}
    \caption{(Color online). 
    The behavior of the excess energy 
    $\Delta E(t)$ after a quench to $T=0$ (left panel) and to $T=1.5$ 
    (right panel) is shown for different system 
    sizes (see key). Averages are taken over $10^3$ realizations. 
    The dashed line is the power law behavior 
    $\Delta E \propto t^{-1/2}$ expected in the scaling regime.
   The area in the box corresponds to the time region where heat exchanges
    are computed (see Sec. (\ref{Heat})).}
\label{figene}
\end{figure}

In Fig. (\ref{figene}) the time 
behavior of the excess energy $\Delta E$ is shown. In the case of a quench to
$T=0$, after an initial transient,
it starts decreasing with the expected power-law behavior up to
a certain time where finite-size effects set in and saturation occurs.
Increasing $N$ the finite-size
effect is delayed, as expected.     
The same behaviour is observed in a quench to $T>0$, the main difference being
a value of $z$ slightly larger than the expected one $z=2$ due to
preasymptotic corrections, as discussed in \cite{isingt0}.

\section{The Ising model in contact with two heat baths} \label{two}

We consider now 
the case of two coupled binary systems 
(denoted simply as 1 and 2)
interacting with thermostats $T_1$ and $T_2$
at different temperatures $T_1>T_2$ (for simplicity we use the same symbols
for the temperature of the reservoirs and the reservoirs
themselves). We denote with
$Q_{T_1,1}$ ($Q_{T_2,2}$) the heat flowing in a time interval
$I_\tau=[t_w,t_w+\tau]$ from the system 1 (2) to the thermostat $T_1$ ($T_2$ ),
and with $Q_{1,2}$ that passing from 1 to 2.

When $T_1, T_2>T_c$, starting from any initial condition, the system 
attains a non-equilibrium stationary state after a certain (microscopic) time. 
The same behavior is observed if one (or both) the temperatures
are below $T_c$, provided that the system(s)
coupled to a sub-critical temperature is prepared in an initial
state with broken symmetry,
in order to avoid the coarsening dynamics. 
The statistics of the heats exchanged between the systems and the thermostats  
has been investigated analytically 
by a mean field approach in \cite{meanf} and numerically in \cite{noi}.
The case where one
or both the systems are interested by phase-ordering kinetics has
not been studied yet, and will be considered in Sec. \ref{Numsetup}.
This case poses the problem of a direct measurement of 
$Q_{1,2}$, as it is discussed below. 

In general the heat exchanged between two parts of a system or between
a part of a system and a thermostat in the
time interval $I_\tau$ can be written as
\be
Q =\sum _{s=t_w}^{t_w+\tau}\,\,\sum _{i=1}^{n(s)} q_i(s)
\label{quantoq}
\ee
where $q_i(s)$ is the heat exchanged by a single degree (a spin, in the Ising model)
in an unit time
around time $s$ and
$n(s)$ is the number of degrees of freedom  which can effectively exchange 
the heat. Eq. (\ref{quantoq}) can be straightforwardly used for the
computation of the heats $Q_{T_1,1}$, $Q_{T_2,2}$. In 
a numerical simulation the $q_i(s)$ are 
the energies $\delta E$ released by the thermostats
in the flip of the $i-th$ spin 
(according to Eq. (\ref{metrop})) in the Montecarlo step occurring at time $s$.

When the system is stationary, a direct computation of $Q_{1,2}$ can be avoided 
because for $\tau $ large enough its value can be related
to $Q_{T_1,1}$ and $Q_{T_2,2}$. Indeed, denoting with 
$\delta E_1^{(\tau)}$ the energy stored in the system 1 
(and analogously for 2) in the time interval $I_\tau$
($\delta E_1^{(\tau)}=E_1(t_w+\tau)-E_1(t_w)$), 
one has $\delta E_1^{(\tau)}=Q_{T_1,1}-Q_{1,2}$ and
$\delta E_2^{(\tau)}=Q_{T_2,2}+Q_{1,2}$. Combining these equations
one has
\be
Q_{1,2}=\frac{1}{2}\left [Q_{T_1,1}-Q_{T_2,2}\right ]+
\frac{1}{2}\left[\delta E_2^{(\tau)}-\delta E_1^{(\tau)}\right].
\label{qdifff}
\ee
In a stationary state 
the quantities $n(t)$ and $q_i(t)$ of Eq. (\ref{quantoq})
fluctuate around a $t$-independent average value. Hence 
$\langle Q_{T_1,1}\rangle$ and $\langle Q_{T_2,2}\rangle$ grow
proportionally to $\tau$. On the other hand the boundary term
$\delta E_2^{(\tau)}-\delta E_1^{(\tau)}$ is finite (for a finite system).
Hence, in the large-$\tau$ limit one can compute $Q_{1,2}$ as
\be
Q_{1,2}\simeq \frac{1}{2}\left [Q_{T_1,1}-Q_{T_2,2}\right].
\label{qdiff}
\ee
Therefore, in this case, it is sufficient to collect the statistics
of the heats $Q_{T_1,1}$ and $Q_{T_2,2}$ exchanged with the thermostats and, from this,
the distribution of $Q_{1,2}$ can be determined through Eq. (\ref{qdiff}),
and Eq. (\ref{fluc}) can be tested. 
This was made in \cite{noi} with a numerical setup where 
a two-dimensional Ising model on a $L$ x $L$ square lattice
was divided into two interacting halves
of size $L/2$ x $L$ (system 1 and 2), in contact with
the two heat baths.

In Sec. \ref{Numsetup} we will prepare the two systems as containing 
fast modes (system 1) and a mixture of fast and slow ones (system 2).
Heat fluxes between different degrees will then be inferred from the knowledge of
$Q_{1,2}$. However, in this case, since 2 is aging, it is no longer true that
$n(t)$ and $q_i(t)$ fluctuate around time-independent values and, hence,
Eq. (\ref{qdiff}) cannot be used. For this reason
we need another operative definition of the quantity $Q_{1,2}$. This is the subject
of the remaining of this section. In order to do that,
it is first useful to understand 
the basic mechanisms whereby heat is transferred at stationarity.
This can be better illustrated by considering first the case 
of two systems made by
one single spin each. We also assume, for the moment, that
the two spins are updated alternately
with Metropolis transition rates.
Denoting with $1_\uparrow $
and $1_\downarrow$ the two possible states of system 1 (and similarly for 2),
let us consider the case of an initial state $1_\uparrow, 2_\uparrow$ which 
evolves in four steps as follows
\be
1_\uparrow, 2_\uparrow   \quad \Rightarrow \quad  
1_\downarrow, 2_\uparrow  \quad  \Rightarrow \quad 
1_\downarrow, 2_\downarrow \quad  \Rightarrow \quad 
1_\uparrow, 2_\downarrow   \quad  \Rightarrow \quad 
1_\uparrow, 2_\uparrow  
\label{scheme} 
\ee
eventually returning to the initial state.
In the initial state the system is in the lowest energy state.
Since 1 is in contact with the higher temperature bath, it has a larger
probability to be flipped first. If this happens, as in the first
step in the scheme (\ref{scheme}), the heat $Q_{T_1,1}=2$ (equal to the
energy increase $\delta E$ of the system) is transferred from
the reservoir $T_1$ to 1. Now it is the turn of 2 to be updated.
With Metropolis rates it flips with probability one, because in doing so the
energy decreases.
Then the heat $Q_{T_2,2}=2$ is released by 2 to $T_2$ (step two).
At this stage a heat $Q=2$ has been transferred between $T_1$ and $T_2$:
Hence we conclude that the amount $Q_{1,2}=Q$ has flown between the two spins.
Now the system is again in the lowest energy state and, as before,
the most probable evolution is the reversal of 1, where a heat $Q_{T_1,1}=2$
is again taken from the reservoir (step three). This move
surely induces the flip of 2 (step four) with a heat $Q_{T_2,2}=-2$
flowing to $T_2$ and another amount $Q_{1,2}=Q$ exchanged between the two
systems. In conclusion, if this process happens, a heat
$Q_{1,2}=4$ flows from 1 to 2. Clearly, one can analogously imagine a process
where $Q_{1,2}=-4$, but this process implies that 2 flips first, which is
less probable. This explains why, on average, $Q_{1,2}$ is positive, even
if negative fluctuations (which in stationary states are regulated by
Eq. (\ref{fluc})) are possible.
The physical mechanism described above is such that heat is transferred from one system
to the other whenever the flip of one spin triggers the flip of the other.
It is then quite natural to introduce a direct measure of the heat flow as 
$Q_{1,2}(\tau)=\sum _{s=t_w}^{t_w+\tau} \delta E(s)\sum _j\epsilon _j\delta _{s,s^{trig}_j}$,
where $\delta E(s)$ is the energy variation of the system in the $s$-th
step belonging to the interval $I_\tau $,
$s^{trig}_j$ is a subset of steps where a flip has been
triggered by a previous reversal of the other spin, and $\epsilon _j=\pm 1$
depending on the direction of the heat flow.  We
state that the flip of (say) 2 in the $s_j$-th step has been triggered if
1 and 2 were not aligned before reversing 2,
and 1 was the last spin flipped before $s_j$. In this case
heat is flowing from 1 to 2 and hence $\epsilon _j=1$. 
In the opposite situation in which 1 is flipping at time $s_j$ and 2
was flipped previously one has $\epsilon _j=-1$.
 
The same definition is 
meaningful if a different sequence of updates
(i.e. random) is assumed and/or different transition rates are considered
(i.e. Glauber). The main difference in this case is that after the flip
of 1, spin 2 is not forced to flip and it might happen that 1
recovers again its original value. However in this case there is
no heat transfer between 1 and 2, and this process does not alter
the computation of $Q_{1,2}$.

Let us now consider the case of systems with many degrees of
freedom. The mechanism described in (\ref{scheme})
is at work between any couple of spins on the interface between 1 and 2.
Clearly, many other processes, involving also
other spins away from the interface, may occur. However, they do not overrule our
algorithm, since there is no heat transfer between 1 and 2 until the process
arrives on the boundary between them. 
Therefore, we can adopt the same operative definition for the 
heat transferred between any couple of spins on the interface, and $Q_{1,2}$ 
is then obtained by summing all these contributions.

In order to test this method, we have prepared a system with the geometry discussed
below Eq. ({\ref{qdiff}) in a stationary state, and we have computed 
$Q_{1,2}$ with this 
technique and, independently, $Q_{T_1,1}$ and $Q_{T_2,2}$. 
If the method is reliable, the determination of $Q_{1,2}$ should
verify Eq. (\ref{qdiff}) for large $\tau$, so the probability 
distributions $P_a(Q)=P(Q_{1,2}=Q)$ and $P_s(Q)=P([Q_{T_1,1}-Q_{T_2,2})/2]=Q)$
should obey $P_a(Q)=P_s(Q)$.
These distributions are compared in Fig. \ref{comparison}. 
For small $\tau$ they are quite different. This is expected due to the presence of 
the boundary term in Eq. (\ref{qdifff}). Increasing $\tau$, however,
the boundary term becomes comparably smaller and the two curves 
tend to merge. 
To be more quantitative we plot, in the inset
of Fig. \ref{comparison}, the {\it distance} 
\be
d_{as}=\sqrt{\int_{-\infty}^{\infty}[P_a(Q)-P_s(Q)]^2 dQ}
\label{distance}
\ee
between the two probability densities, showing that
it is a steadily decreasing function.
This shows that the algorithm for the direct measurement of
$Q_{1,2}$ is reliable. 
When one (or both) of the systems is aging, Eq. (\ref{qdiff})
is spoiled by the extra heat released to the baths because
the system is lowering its energy due to relaxation.
In these cases the direct computation of $Q_{1,2}$ according to
the technique described above is mandatory.    
In the following, it will always be used to compute $Q_{1,2}$.
 
\begin{figure}
    \centering
    
   \rotatebox{0}{\resizebox{.6\textwidth}{!}{\includegraphics{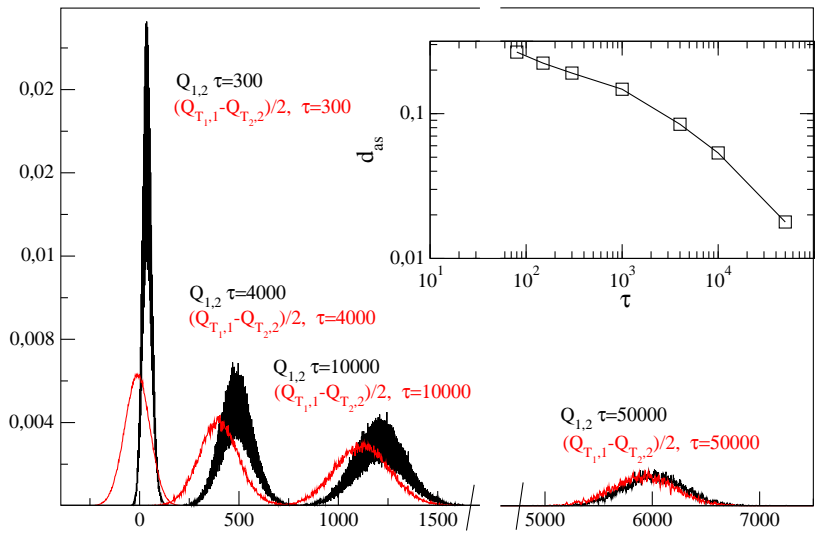}}}
\vspace{3cm}
    \caption{(Color online).Comparison between $P[(Q_{T_1,1}-Q_{T_2,2})/2]$ 
and $P(Q_{1,2})$ computed in a system of size $L=50$ in a stationary state 
with $T_1=2$ and $T_2=1.5$,
for different values of $\tau$, as specified in the plot. 
The statistics is collected over
$1.7 \times 10^6$ realizations. In the inset the {\it distance} $d_{as}$ between
the two distributions is plotted.}
\label{comparison}
\end{figure}

\section{Heat exchanges in phase-ordering} \label{Numsetup}

In this section we study the heat exchanges in phase-ordering systems.
We do this with the help of some numerical experiments designed to
collect the statistics of the heat
flowing between fast and slow degrees and among them and the
reservoir. This will allow us to test 
the scenario with two statistically independent degrees coexisting at
different (effective) temperatures and to discuss why such independence
may be retained in the evolution. 

In order to do that, the straightforward procedure would be
to consider an aging system, to recognize and discriminate (at any time)
fast and slow components, and hence to detect the heat they are exchanging.
However, this is not easily realizable. Indeed, although in a coarsening
system we have an idea of what slow and fast modes are, 
a precise definition and separation of degrees is to a large
extent arbitrary. Moreover, if even one could find a
technique suited to do that, this would be tuned to phase-ordering
and the method would be restricted to coarsening systems.
Instead, we propose a different approach where two
systems are coupled. Preparing such systems so that the first is in 
equilibrium while the second coarsens, as will be detailed below,
one may assume that slow degrees are confined in the second.
As we shall discuss, this will allow us to proceed to the discussion
of heat exchanges.  
In doing that, we adopt the same geometry as
for the stationary case described in the previous section except that, 
in order to guarantee that the heat exchanged is small enough we have
activated only two links, at distance $L/2$, between the two systems.
In all the cases we have chosen the parameters
in such a way that the system is free from finite-size effects.
System 1 is prepared in equilibrium at the temperature $T_1$
while system 2 is in an infinite temperature disordered configuration. 
At time $t=0$, sample
2 is brought in contact with the thermostat at the temperature $T_2<T_c$ while
1 interacts  with a bath at the same temperature $T_1$ of its initial 
equilibrium state. 
Then, from time $t$ onwards 2 ages via phase-ordering kinetics, 
while 1 remains in a stationary configuration.
Precisely, 1 is not stationary in principle, due to the
coupling with the non-stationary system 2, but this effect is negligible
for large system sizes and few active links between the two sub-systems. Moreover
the effect is further suppressed in the asymptotic time domain, 
due to the progressive decrease of the number of slow degrees in sample 2. 
Indeed our simulations did
not show any deviation from stationarity in 1. 
The statistics of the heats is collected from some time $t_w$ to $t_w+\tau$
in a sample of size $L=100$.
Notice that in the range of times considered in this simulation, represented 
by the dashed box in the right panel of
Fig. \ref{figene}, finite-size effects can be neglected.

Indicating with $F_1$, $F_2$ the fast degrees of freedom
of the two coupled systems, with 
$S_2$ the slow ones, and with $Q_{F_1,F_2}$
the heat exchanged between $F_1$ and $F_2$ (and similarly
for the heat exchanged between fast and slow ones), 
the general scheme of all the possible heat exchanges is summarized in
figure \ref{schema}.

\begin{figure}
    \centering
    
   \rotatebox{0}{\resizebox{.6\textwidth}{!}{\includegraphics{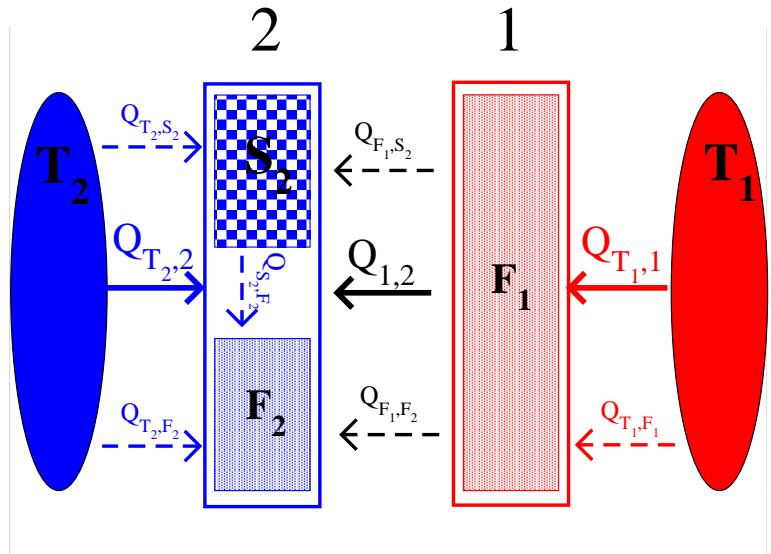}}}
\vspace{1cm}
    \caption{(Color online). Schematic description of the system according to
the two-degrees scenario. Systems 1 and 2 are represented
as blue and red boxes, and are in contact with the reservoirs $T_1$ and $T_2$,
represented as ellipses. The heat exchanged between the two systems ($Q_{1,2}$),
and between the systems and the respective reservoirs ($Q_{T_1,1}$ and $Q_{T_2,2}$)
are represented as bold arrows (heats flowing in the direction of the arrows are
considered positive). Fast and slow degrees are represented 
as separate subsystems by the boxes $F_1$, $F_2$ and $S_2$.
All the possible heat exchanges among them and between them and the
reservoirs are indicated with dashed arrows.}
\label{schema}
\end{figure} 

We start with the case in which the two baths are at the same
temperature $T_1=T_2=1.5$. This will allow us to draw some conclusion
on the possibility that, in a single coarsening system, heat is
released directly from the slow modes to the fast ones.
The probability distributions of  $Q_{T_1,1}$, $Q_{T_2,2}$ 
and of $Q_{1,2}$ have been computed letting $x=\tau/t_w$ fixed while
$t_w$ is increased (we used values of $t_w$ up to
$t_w=1000$, for different choices of $x$).
This will be referred to as the aging limit.
For every choice of $t_w$ and $x$ we found a pattern
of behaviors analogous to the one shown  
in Fig. \ref{A1},  where $t_w=80$ and $x=1$. 
Here one sees that $\langle Q_{T_2,2}\rangle  < 0$, because 2 is 
transferring the excess energy
associated to the interfaces to the bath $T_2$.
On the other hand the heat $\langle Q_{1,2}\rangle $ is very small.
This happens because with the present geometry 
$Q_{1,2}$ can be exchanged only through two links and
because we are working in the aging limit.
Indeed, while $F_1$ and $F_2$ are equilibrated at the same temperature, and hence 
$\langle Q_{F_1,F_2}\rangle =0$, $\langle Q_{F_1,S_2}\rangle  < 0$
is the only contribution to $\langle Q_{1,2}\rangle$.
However, in the aging limit
since $\tau \propto t_w$ and the number of slow modes which contribute
to $\langle Q_{1,2}\rangle $
is $\langle n(t_w)\rangle\propto t_w^{-1/2}$, 
there is a number of order $t_w^{1/2}$ of terms in 
the double sum in Eq. (\ref{quantoq}).
On the other hand, for the heats exchanged involving only fast degrees, such as
$\langle Q_{F_1,F_2}\rangle$, $\langle Q_{F_1,T_1}\rangle$,
and $\langle Q_{F_2,T_2}\rangle$, $n(t_w)$ is constant and hence
there is a number of order $t_w$ of terms contributing 
in Eq. (\ref{quantoq}). 
Therefore, the heat $\langle Q_{1,2}\rangle $ is negligible with respect to 
the others.
This shows that our experiments
are performed in the limit of small average heat exchanged between the
two systems. However, although small,
$\langle Q_{1,2}\rangle $ is not strictly zero and its behavior parallels 
that of $Q_{T_2,2}$. This can be seen in Fig. \ref{figQ12vsQ2}
where $\langle Q_{1,2}\rangle$, $\langle Q_{T_2,2}\rangle$ and their ratio
is plotted for different choices of $x$.
In the inset of this figure one sees that, after an initial transient,
$\langle Q_{T_2,2}\rangle$ decays as $t_w^{-1/2}$. A similar behavior is 
found for $\langle Q_{1,2} \rangle$,
the main difference being an oscillating behavior superimposed on the
$t_w^{-1/2}$ decay.
The origin of such oscillations is
not completely clear and is probably related to the recurrent
passage of interfaces across the boundary links.
Their amplitude can be suppressed by increasing $x$,
because collecting the heats on larger time windows
effectively averages out the oscillating behavior.
For $x\gtrsim 50$ the periodic behavior is basically washed away.
Notice also that the curves tend to superimpose
as $x$ increases.
The ratio $\langle Q_{1,2}\rangle/\langle Q_{T_2,2}\rangle$
is plotted in the main part of the figure, showing that
for sufficiently large values of $t_w$ and $x$ it converges to 
a constant value, namely $\langle Q_{1,2}\rangle$ and $\langle Q_{T_2,2}\rangle$
are proportional.
This behavior supports the hypothesis that
slow degrees transfer heat to the fast ones and to the reservoir
basically in the same way. 
This is so because, since fast degrees
are thermalized with the reservoirs they affect the slow ones 
similarly to what the
thermal baths do, namely absorbing heat.  

\begin{figure}
\vspace{0.3cm}
\rotatebox{0}{\resizebox{.5\textwidth}{!}{\includegraphics{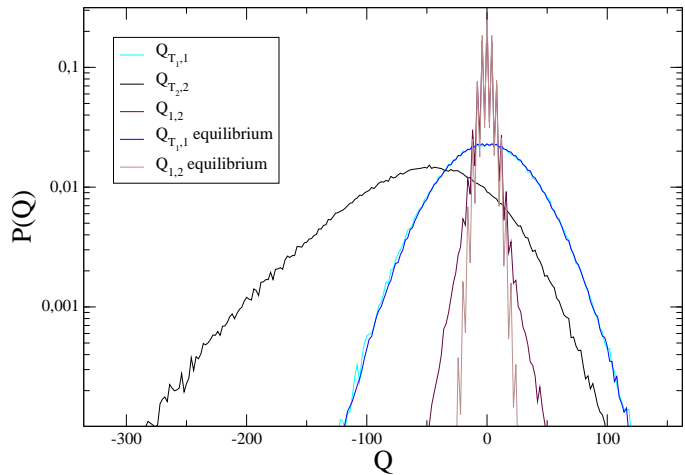}}}
\vspace{1cm}
\caption{(color online). $P(Q_{T_1,1})$, $P(Q_{T_2,2})$ and $P(Q_{1,2})$ for an aging system coupled
to a (quasi) stationary one in contact with baths at the same temperatures
$T_1=T_2=1.5$, with $t_w=80$ and $x=\tau/t_w=1$.
The distributions $P(Q_{T_1,1})$ and $P(Q_{1,2})$ of 
the case where both sub-systems are in equilibrium at the same temperature 
$T_1=T_2=1.5$ are also shown. The statistics is collected over $2\times 10^5$ realizations.}

\vspace{1cm}

\label{A1}
\end{figure}

\begin{figure}
\vspace{0.3cm}
\rotatebox{0}{\resizebox{.5\textwidth}{!}{\includegraphics{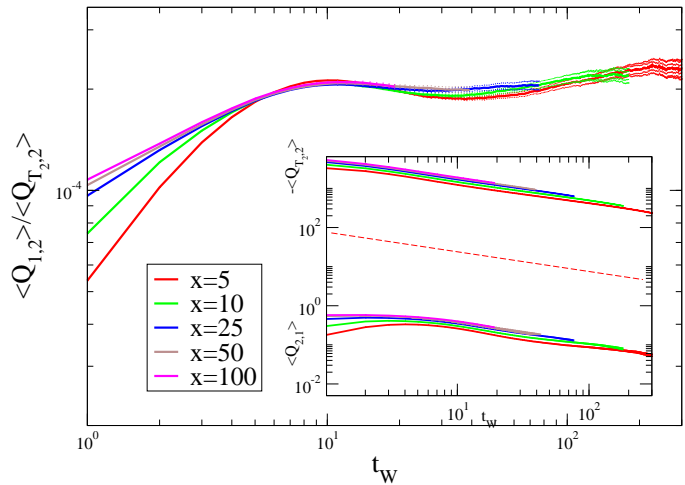}}}
\vspace{1cm}
\caption{(color online). The ratio
$\langle Q_{1,2}\rangle/\langle Q_{T_2,2}\rangle$ is plotted against $t_w$ for
different values of $x$ (see key). Data are averaged over $3.6 \times 10^7$ realizations.
The clouds around the bold curves represent
the error bars. In the inset $\langle Q_{1,2}\rangle$ and $\langle Q_{T_2,2}\rangle$ 
are plotted separately. The dashed line is the power law $t_w^{-1/2}$.}
\label{figQ12vsQ2}
\vspace{1cm}
\end{figure}

Let us see how these findings fit into
the scenario with two degrees coexisting at different
(effective) temperatures. This framework implies that statistical independence
is retained and thermalization does
not occur between them. The question is why. 
The most obvious explanation would be that slow and fast degrees
do not effectively interact; they are {\it insulated}, in some sense.
However, the above experiment has shown that this is not the case,
since heat is transferred from slow degrees to the fast ones and to the
reservoir basically in the same way.
Hence, statistical independence 
must have a different origin.
A possible explanation is the following:
Let us consider two separated domains at time $t_1$ (left part of 
Fig. (\ref{due_interf})). They successively merge at 
the later time $t_2>t_1$, releasing some heat $Q$ to the fast
degrees surrounding the coalescence event. This excess heat will
outflow to the bath on the microscopic time $t_{eq}$.  
After the event
the slow degrees which have originally released the heat 
are annihilated. The surviving ones 
are left unmodified by this process, and hence retain their original
properties. In summary, if energy release is always accompanied by annihilation, 
the heat transferred between fast and slow components up to a certain time 
was provided by interfaces that do not exist any more, while those still surviving
have never been involved. 
Notice that, writing $\Delta E(t)=\Delta E(t_w)+Q(\tau)$,
where $Q(\tau )$ is the heat released to the reservoir, 
from Eq. (\ref{eq:E}) one has $Q(\tau )\propto n_s(t)-n_s(t_w)$, where
$n_s$ is the number of slow degrees. This shows that $Q(\tau)$
is proportional to the decrease of $n_s$, and this is consistent with the
hypotheses that heat release and annihilation are deeply
related processes.

\begin{figure}
    \centering
    
   \rotatebox{0}{\resizebox{.6\textwidth}{!}{\includegraphics{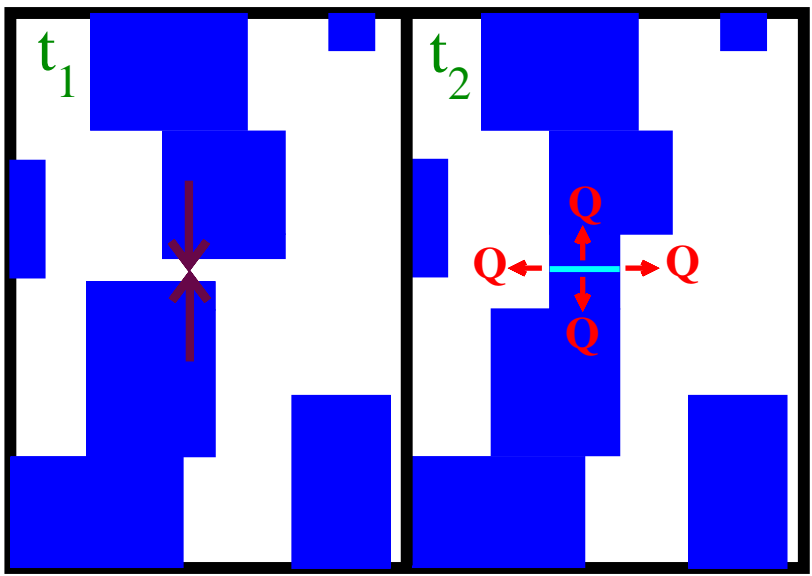}}}
\vspace{1cm}
    \caption{(Color online). Schematic description of the process
of interface annihilation and subsequent heat release to the fast degrees.}
\label{due_interf}
\end{figure} 

Having shown 
that fast degrees absorb the excess heat released by 
the interfaces 
it remains to be shown that 
this process does not spoil
the thermal properties of the fast degrees, preserving in this way 
the statistical independence. This is expected  
because bulk degrees are able to thermalize among themselves and with
the reservoirs in a microscopic time $t_{eq}\ll t_s$, making 
their thermal properties irrespective of the presence of interfaces. 
If this is true, one might predict that 
when two coarsening systems quenched to different temperatures $T_1$, $T_2<T_1$
are put in contact, heat flows from the fast components of the hotter system to
those of the colder. Moreover, fluctuations of heat
exchanged between fast modes should be similar to those 
observed in stationary systems and, in particular, governed by the 
fluctuation principle (\ref{fluc}).}
Strictly speaking, we should verify Eq. (\ref{fluc}) but with
$Q_{F_1,F_2}$ in place of $Q_{1,2}$. However,
we recall that Eq. (\ref{fluc}) is supposed to hold in the large-$\tau$
domain where boundary terms can be neglected, similarly to
what happens in Eq. (\ref{qdifff}).
In this regime the number of bulk spins is much larger than that of the
interfacial ones and so $Q_{F_1,S_2}$ is negligible with respect to
$Q_{F_1,F_2}$. Then, if the thermal properties of fast degrees is
unaffected by the presence of the slow ones one should be able 
to verify the FR (\ref{fluc}) for the {\it whole} heat transferred $Q_{1,2}$.

\begin{figure}
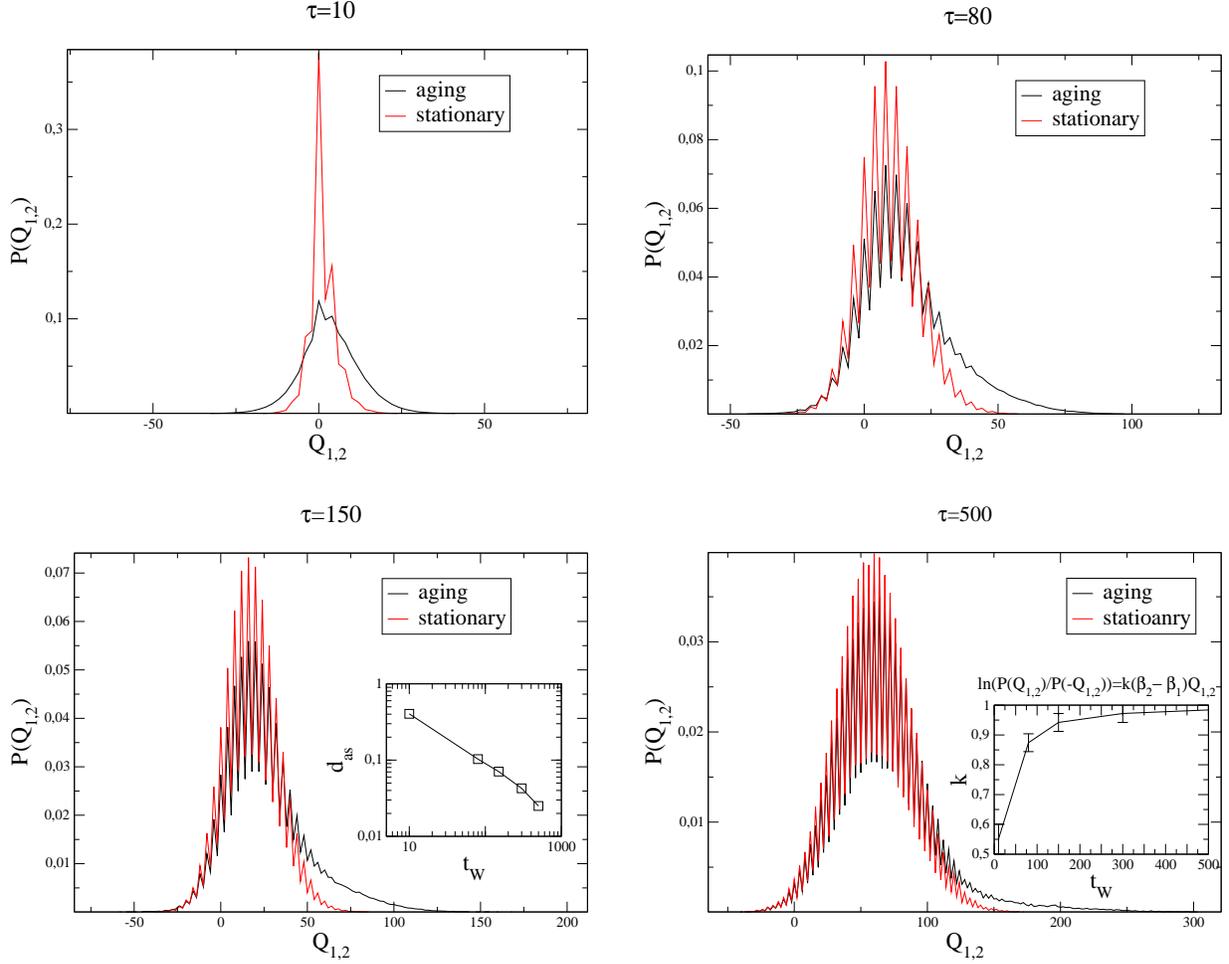

\vspace{0.3cm}
\includegraphics[width=0.43\textwidth]{Q12_aging_stationary_tw10_tau10.eps}
\hspace{.5cm}
\includegraphics[width=0.43\textwidth]{Q12_aging_stationary_tw80_tau80.eps}
\\
\vspace{.5cm}
\includegraphics[width=0.43\textwidth]{Q12_aging_stationary_tw150_tau150.eps}
\hspace{.5cm}
\includegraphics[width=0.43\textwidth]{Q12_aging_stationary_tw500_tau500.eps}
\caption{(Color online). Comparison between $P(Q_{1,2})$ computed in a stationary state
with $T_1=2$, $T_2=1.5$ and in the setup where 2 is (quasi)-stationary
at $T_2=1.5$ while 1 is aging after a quench to $T_1=2$. 
The system size is $L=50$. 
Different
values of $\tau$ are shown, with $x=\tau/t_w=1$.
The statistics is collected over $10^6$ realizations.
In the inset of the left-bottom panel the {\it distance} $d_{as}$ is
plotted against $t_w$. In the inset of the right-bottom panel
the values of the slope $k$ obtained from the
fluctuation-relation (see text) are plotted against $t_w$.}
\label{A3}
\end{figure}

In order to study this,
we have designed an experiment similar to the one discussed above. However
now we quench system 1 to $T=T_1=2<T_c$, while 2 is in equilibrium at
$T_2=1.5$.
In Fig. \ref{A3} we compare the probability distribution $P_a(Q_{1,2})$ of the
heat $Q_{1,2}$ in this experiment, with the one  $P_s(Q_{1,2})$ measured 
in the experiment described in Sec. (\ref{two}) where the samples are
again in contact with the same baths at $T_1=2, T_2=1.5$, but they are both 
stationary. 
The two distributions have a zig-zag shape. This is not due to a lack of
statistics but is a genuine feature that we will discuss later.
Notice that, at any finite time, the positive tail of $P_a(Q_{1,2})$
is fatter than that of $P_s(Q_{1,2})$. 
This is due to the extra heat flowing from the coarsening system 1 to
the stationary sample 2, which is produced by the reduction of the interface density,
similarly to the case with $T_1=T_2$ discussed above (see Fig. \ref{A1}).
Besides this, one observes that the two distributions tend to merge in
the large time limit when all boundary terms become negligible
and slow degrees are sufficiently few. 
We made this observation more quantitative by plotting, in one inset
of Fig. \ref{A1}, the {\it distance} 
$d_{as}$
between the two probability densities, defined in Eq. (\ref{distance}), 
which decreases in a power-law way.
The merging of the two distributions indicates
that the thermal properties of the fast degrees
are left unchanged by the presence of the slow ones.
As a further check we consider the validity
of Eq. (\ref{fluc}): 
After having verified that the l.h.s. of Eq. (\ref{fluc}) is well fitted
by $\ln [P(Q_{1,2})/P(-Q_{1,2})]=k(\tau)(1/T_2-1/T_1)Q_{1,2}$ 
for all the values of $\tau$, we have plotted the {\it slope}  
$k(\tau)$ in the inset of the bottom-right panel of Fig. \ref{A3}.
One sees that $k(\tau)$ converges to $k(\tau)=1$
when $\tau $ is increased, as expected. 

Let us now come back to the zig-zag feature of the distributions of Fig. \ref{A3}.
which can be explained as follows.
Let us consider the 4-steps microscopic heat transfer process 
(\ref{scheme}).  According
to our operative definition of $Q_{1,2}$, one measures
$Q_{1,2}=2$ or $Q_{1,2}=4$ after the first or after the fourth step,
respectively. This implies that $Q_{1,2}$ is always a multiple of
2, in this model.
In the situations considered in this Section spin 2 is part
of an extended system in contact with the reservoir at $T_2<T_c$. 
This system is in a coarsening stage, where the up-down symmetry
is not globally broken. However locally, the neighborhood of spin 2 has almost
surely a
broken symmetry character, if time is large enough. 
Indeed, this spin will almost
surely be located in the bulk of a magnetized domain. 
Suppose
that symmetry is broken around the $\uparrow$ direction. 
In this situation the probability of finding the first or the last
configuration of the process (\ref{scheme}) is much higher
than that of finding the other. 
Since the final time $t=t_w+\tau$ over which the heats
are collected is not correlated to the spin configurations,
one has a much higher probability that $Q_{1,2}$ is a multiple
of 4 than that it is not. This explains the zig-zag behavior
where values of $Q_{1,2}$ which are multiples of 4 are enhanced.

\section{Conclusions and perspectives} \label{conclusions}

Coarsening systems are
prototypical models which exhibit a non trivial aging behavior 
where two classes of degrees of freedom with different properties
can be identified and studied to some extent. They are then
particularly suited to understand the issue of
thermal exchanges in aging states and the related concept
of effective temperatures. In this perspective, in this
paper we have studied numerically the heat flows 
occurring in an Ising model quenched below the critical temperature.
In order to do that we have first developed an algorithm to quantify
the heat transferred between two parts of a system, and we have subsequently used 
this tool in a series of numerical experiments designed as to
detect the heat fluxes occurring between the two kinds of degrees
existing in the coarsening system. Our results fit into a scenario where
fast components thermalize with the reservoirs and act themselves
as baths where interfaces can temporary store their excess energy
when annihilation events occur. The mechanism is such that, after 
a microscopic time of order $t_{eq}$, the statistical properties of
both fast and slow modes are left unchanged. This guarantees that
the two kind of degrees may remain statistically independent,
as shown analytically in a soluble limit in \cite{noilargen}.
Once a thermodynamic interpretation of the effective temperature
(in the spirit of Ref. \cite{peliti}) is agreed upon
(let us mention however that our results do not speak about the 
consistent interpretation of $T_{eff}$ as a genuine 
thermodynamic temperature) our results help to explain
why a couple of different effective temperature can be sustained
indefinitely. At variance to what is naively expected, this does not 
occur because the two classes of degrees are thermally insulated,
or not effectively interacting. On the contrary, heat is exchanged
between them much in the same way as it is exchanged with the baths.
The mechanism whereby this occurs, nevertheless, is such to
preserve the statistical properties of the two kind of modes.
We have given support to this allegation by proving that
the Gallavotti-Cohen FR (\ref{fluc}) is obeyed in the asymptotic domain
also when aging occurs in systems at different temperatures. 
Let us stress here that 
$T_1$ and $T_2$ entering Eq. (\ref{fluc}) are in this case the temperatures
of the thermal baths. The presence of $T_{eff}$ does not play
an appreciable role in our experiments because slow degrees do not contribute significantly
to the heat exchanged $Q_{1,2}$, which is dominated by the contribution
of the fast spins. Interestingly, $T_{eff}$ might be
expected to enter the FR \cite{cugzam} if only the heat exchanged by
slow degrees was considered,  (namely subtracting away the dominant contribution
provided by the fast ones in a specially designed experiment).

\vspace{1cm}
\noindent
\underline{Acknowledgements.}

F.C. acknowledges financial support from PRIN 2007JHLPEZ 
({\it Statistical Physics of Strongly correlated systems in Equilibrium
and out of Equilibrium: Exact Results and Field Theory methods}).

\end{document}